\documentstyle[aps,epsfig,floats,prb]{revtex}

\begin{document}
\draft
\title{Quantum phase transitions in the
Triangular-lattice Bilayer Heisenberg Model}

\author{Rajiv R. P. Singh}
\address{Department of Physics, University of California, 
Davis, California 95616}
\author{Norbert Elstner}
\address{Physikalisches Institut, Universit\"at Bonn, 
         Nu\ss allee 12, D-53115 Bonn, Germany }

\twocolumn[\hsize\textwidth\columnwidth\hsize\csname
@twocolumnfalse\endcsname
\date{\today}
\maketitle
\widetext
\begin{abstract}
  \begin {center}
    \parbox{6in}{ We study the triangular-lattice bilayer Heisenberg
  model with antiferromagnetic interplane coupling $J_\perp$ and
  nearest-neighbor intraplane coupling $J=\lambda J_\perp$, 
  which could be ferro- or antiferromagnetic, by expansions in $\lambda$.
  For negative $\lambda$ a phase transition is found to an
  ordered phase at a $\lambda_c= -0.2636\pm 0.0001$, which is in the
  $3D$ classical Heisenberg universality class.
  For $\lambda>0$, we find a transition at a rather large
  $\lambda_c\approx 1.2$. The universality class of the transition
  is consistent with that of Kawamura's $3D$ antiferromagnetic stacked
  triangular lattice. The spectral weight for the triplet excitations,
  at the ordering wavevector, remains finite at the transition, suggesting
  that a phase with with free spinons does not exist in this model.}
  
  \end{center}
\end{abstract}

\pacs{\hspace{0.5in}PACS: 75.10.Jm, 75.40.Cx, 75.40.Mg, 75.50.-y, 75.50.Ee}

]

\narrowtext

In recent years much interest has focussed on the nature of quantum
disordered phases of the Heisenberg antiferromagnets, where the
combination of low dimensionality, low spin and frustration
cause the ground state of the system to be disordered.\cite{CHN,CSY} The case
of one-spatial dimension is relatively well studied and
understood. Two dimensional systems have received particularly
large attention, due to their relevance to high
temperature superconductivity. However, despite much effort,
quantum disordered phases are not fully understood in $d=2$.

A special situation are those types of quantum disordered phases where
the ground state is to a good approximation a product of local
singlets over even-spin clusters. This can arise due to explicitely
dimerized (or clustered) Hamiltonians; a situation that appears to be
relevant for the material CaV$_4$O$_9$.\cite{CAVO} Another scenario
is the spontaneous breaking of translational symmetery, as found in
large-$N$ theories and also suspected in several frustrated
models, which leads to dimerization.\cite{LargeN,GSH,ZimanSchulz}
In all these systems, the elementary excitations are triplets with a
finite excitation energy.

In contrast to these, a different class of quantum disordered
phases would be one where the elementary excitations are 
free spin-half objects or spinons. Such phases, for $d=2$, have
been predicted in systems where the classical ground state
is non-colinear\cite{CSS} and their properties have been investigated
by field-theoretic methods\cite{Starykh}. However, no lattice models
are known where such a behavior is realized. One potential
candidate system for such a behavior is the Kagome-lattice
antiferromagnet, where the ground state is widely believed to be
magnetically disordered.\cite{kagome0,SH}

Here we study the triangular-lattice bilayer Heisenberg model. The
model consists of two-layers of triangular lattices, one on top of the
other, with an intralayer nearest neighbor Heisenberg coupling $J$,
which could be ferro- or antiferromagnetic, and an antiferromagnetic
interlayer nearest neighbor coupling $J_\perp$, between the spins on
top of each other.  This model maybe relevant to bilayer quantum-hall
systems\cite{SachdevFQHE} and to layers of He$^3$.

The corresponding square-lattice Heisenberg bilayer has been
extensively studied by many authors.\cite{Square} 
It was found that with increasing
$\lambda$ the triplet excitations at the ordering wavevector soften
and at a critical $\lambda\approx 0.40$ 
the gap closes and there is a
continuous quantum phase transition to a magnetically ordered
phase. It should be mentioned that the same scenario is found for
ferromagnetic intraplane coupling $\lambda <0$ with a critical point
at $\lambda_c\approx -0.44$, which, to our knowledge, has not been
reported in the literature before.

We study the triangular-bilayer model by a strong coupling expansion
in the parameter $\lambda=J/J_\perp$. At $\lambda=0$, the ground state
consists of products of singlets over pairs of spins and the
elementary excitations are isolated triplets.  We focus on the 
ordering susceptibility, the triplet
excitation spectrum and its spectral weight as a function of $\lambda$.
For ferromagnetic intralayer couplings, the model exhibits a
continuous quantum phase transition to an ordered phase at a critical
coupling $\lambda_c\approx -0.2636$. Our results for the critical
exponents associated with the closing of the triplet gap and the
divergence of the magnetic susceptibility are consistent with those of
the $3D$ classical Heisenberg model.

For the antiferromagnetic intralayer coupling, it is more difficult to
reliably estimate the critical $\lambda_c$, at which the gap vanishes,
as $\lambda_c$ is large and the convergence of the series not very
good at these values of $\lambda$. 
Unbiassed analysis for the susceptibility and the inverse of the
gap series shows evidence for a transition
at $\lambda_c \approx 1.2$ 
with an exponent $\nu\approx 0.53$ and $\gamma\approx 1.1$.
These results are also confirmed by biassing the critical point
in the series analysis. These results
are consistent with the universality class
discussed by Kawamura\cite{kawamura} for the stacked triangular 
Heisenberg antiferromagnet.

These results also appear to rule out a
phase in the model, at intermediate values of $\lambda$, 
where there is still a gap in the spectrum but
unbound spinons become the elementary excitations.
It has been shown\cite{CSS,Starykh} that
in this case the order-disorder transition will lie
in the universality class of
the $3$-dimensional $O(4)$ model. One might also expect that
in this case the triplet excitations would decay into the two-spinon
continuum and not have a well defined energy momentum relation.
However, we find that as $\lambda$ is increased in the model
and the minimum of the triplet spectrum becomes more
pronounced, the spectral weight of
the triplets is reduced over much of the Brillouin
zone but it stays finite and grows with $\lambda$ in the vicinity of the
ordering wavevector. Together with the estimates for
the critical exponents at the transition, 
this suggests that free spinons
do not exist in this model.
Our results provide further support for existence of antiferromagnetic
order in the single-plane triangular-lattice 
antiferromagnet.\cite{SH,triangle0,ESY}

The triangular-lattice bilayer Heisenberg model is given by
the Hamiltonian:
\begin{equation}
   \label{bilayer}
   {\cal H} = J_\perp \sum_{i} {\bf S}_{A,i} \cdot {\bf S}_{B,i} 
            + J \sum_{<i,j>} [{\bf S}_{A,i} \cdot {\bf S}_{A,j} 
                                +{\bf S}_{B,i} \cdot {\bf S}_{B,j}] \;\;, 
\end{equation}
where $A$ and $B$ refer to the two layers of the triangular lattice and
$<i,j>$ are nearest neighbours in a given layer of the lattice. 
The triangular lattice sites are spanned by
the two nonorthogonal primitive vectors
$$
{\bf e_1} = (1,0) \hskip 36pt {\rm and} \hskip 36pt 
{\bf e_2} = \frac{1}{2}(-1,\sqrt{3})
$$

For $J=0$, spins are coupled only in pairs and the ground state
consists of product of singlets over these pairs. The excitations are
isolated triplets localised at some site $i$.  For finite values of
$\lambda=J/J_\perp$ an effective hamiltonian ${\cal H}^{\rm eff}({\bf
R}_{i,j})$ describing the interaction between these localised
degenerate triplet states can be derived by a systematic expansion in
$\lambda$:
$$ {\cal H}^{\rm eff}({\bf R}) =\sum_n \lambda^n h_n({\bf R})$$
  The methods for calculating  ${\cal H}^{\rm eff}$ in
powers of $\lambda=J/J_\perp$ are well developed and discussed in the 
literature.\cite{Gelfand96}  The excitation spectrum is given by the
eigenvalues of the effective Hamiltonian ${\cal H}({\bf R})$, where
${\bf R}_{i,j} = {\bf r}_i - {\bf r}_j$ is the vector connecting sites 
$i$ and $j$. It can easily be diagonalised by a Fourier transform.
We have calculated these quantities complete to $10$-th order.

In table I, the expansion coefficients for
$E({\bf q})$ with ${\bf q} =0$ and
${\bf Q}_{\rm AF} = 4\pi/3 \; {\bf e_1}$ are presented, corresponding to
the ordering wavevectors for the ferromagnetic and the antiferromagnetic
systems. The expansion coefficients for the magnetic susceptibilities
at the same two wavevectors,\cite{Square,method} calculated to $10$-th order
are given in the table II.

\begin{table}
\label{table:Eq}

\begin{tabular}{ | r | r | r | }
             &  ${\bf q}=0$\phantom{WW} & ${\bf q}={\bf Q}_{\rm AF}$\phantom{WW} \\
\hline 
           0 &  1.0              &   1.0             \\
           1 &  3.0              & -1.5              \\
           2 & -1.5              &  1.875            \\
           3 & -0.75             & -1.875            \\     
           4 & -4.125            &  0.4453125        \\
           5 &  15.6328125       & -1.48828125       \\
           6 & -26.9443359375    &  14.68505859375   \\
           7 &  56.977294921875  & -43.2284545898437 \\
           8 & -244.898391723633 &  84.7458343505859 \\
           9 &  794.878423690796 & -241.322682380676 \\   
          10 & -2301.24252033234 &  911.485698580742 \\
\hline 
\end{tabular}  

\caption{Series for the triplet energies $E(\bf{q})$}

\end{table}

\begin{table}
\label{table:chiq}

\begin{tabular}{ | r | r | r | }
             &  ${\bf q}=0$\phantom{WW} & ${\bf q}={\bf Q}_{\rm AF}$\phantom{WW} \\
\hline 
           0 &  0.25000000000000  &  0.25000000000000  \\
           1 &  -1.5000000000000  &  0.75000000000000  \\
           2 &   7.1250000000000  &  0.93750000000000  \\
           3 &  -30.750000000000  &  0.60937500000000  \\
           4 &   128.47656250174  &  -0.1015624982634  \\
           5 &  -528.79947916114  &  1.9749349013237   \\
           6 &   2148.0511067806  &  0.24088542676474  \\
           7 &  -8631.9544994009  &  -1.0813931891307  \\
           8 &   34434.857183080  &  -13.218768766815  \\
           9 &  -136648.82476477  &   59.415134251052  \\
          10 &   539861.35304263  &  -93.997993646134  \\
\hline 
\end{tabular}  

\caption{Susceptibility series $\chi(\bf{q})$}

\end{table}

In addition to the wavevector dependent susceptibilities and the
excitation spectra, we also calculate series for the spectral
weights associated with the excitations.
The spectral weights are defined
by the delta-function piece of the dynamical correlation function.
$$ S({\bf q},\omega)=A({\bf q})
\delta(\epsilon({\bf q})-\omega)+B({\bf q},\omega)$$
They are calcualted via the spin-spin correlation functions,
where the intermediate states are restricted to the elementary
triplet excitations.\cite{SG} These latter calculations
are more difficult and are only done to $6$th order.

For ferromagnetic intraplane couplings, the critical point occurs at
small values of $\lambda < 0$, so even with relatively short series we
can determine the critical point quite well and also get reasonable
estimates for the critical exponents. For the susceptibility series,
the dlog Pade approximants lead to estimates
$$\lambda_c=-0.26362\pm0.00009,\qquad\gamma=1.407\pm0.004.$$ 
Here the uncertainties reflect the spread between different Pade estimates.
Applying d-log Pade approximants to the inverse of the energy-gap
series one obtains, 
$$\lambda_c=-0.2641\pm 0.0005, \qquad\nu=0.73\pm0.01.$$ 
Here, the estimates for the critical points and exponents from
individual approximants are correlated and the more negative the
critical point estimates, the larger is the exponent. Given the length
of our series, these numbers are in quite good agreement with the $3d$
classical Heisenberg university class, where the best current estimates
come from field-theory
$\gamma = 1.3866 \pm 0.0012$, $\nu = 0.7054 \pm 0.0011$.\cite{3dheisenberg}
That the series analysis gives slightly higher estimates for the exponents
is common to many models and is primarily due to corrections to
scaling.\cite{liu-fisher}

We now consider the analysis for $\lambda>0$, which corresponds to
antiferromagnetic intraplane couplings.
We analyzed the inverse of the energy-gap and the
ordering susceptibility series using d-log Pade approximants and 
differential approximants. In this case, the convergence was much
poorer as the critical point occurs at much larger $\lambda_c$.
The estimates for the critical points and the exponents show
tremendous scatter. Assuming that the two series have the same
critical point, the most consistent estimate for
$\lambda_c$ is in the range $1.18-1.29$. In that range there
are four approximants for the susceptibility series, which give
($\lambda_c,\gamma$) values of (1.19,1.10),(1.19,1.08),
(1.21,1.06),(1.26,1.32) and four approximants for the inverse
gap series, which lead to ($\lambda_c,\nu$) values of
(1.18,0.45),(1.22,0.51),(1.25,0.59),(1.29,0.57).
These lead us to conclude that $\lambda_c\approx 1.2$ and that
$\nu \approx 0.53$ and $\gamma\approx 1.1$.
These exponents are also obtained if the approximants
are biassed to have the critical point near $\lambda_c=1.2$.
These results are consistent with
Kawamura's universality class for the stacked
triangular-lattice Heisenberg model, where he found $\nu\approx 0.55$,
and $\gamma\approx 1.1$,\cite{kawamura} and
not consistent with the $O(4)$ universality class, which
has $\nu\approx 0.74$ and $\gamma\approx 1.47$.\cite{o4exp} 
Following Chubukov, Sachdev and Senthil,\cite{CSS} such a
behavior is to be expected if there is a direct transition from the
disordered phase without free spinons to the $3$-sublattice ordered
phase.

\begin{figure}
  \protect\centerline{\epsfig{file=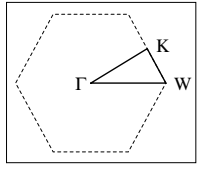, height=1.5in}}
  \caption{Brillouin zone of the triangular lattice}
\label{fig:brillouin}
\end{figure}

\begin{figure}
  \protect\centerline{\epsfig{file=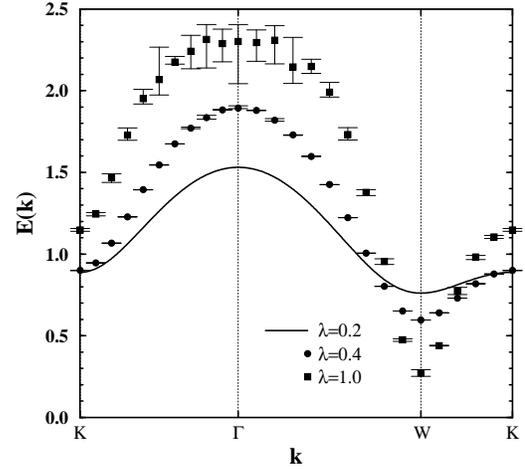, height=2.5in}}
  \caption{Dispersion of the triplet excitations}
\label{fig:disp}
\end{figure}

Another way to explore the existence of an intermediate phase
with free spinons is to study the spectral weights for the
triplets and see if it vanishes as $\lambda$ is increased. 
When the spinons become the elementary excitations,
the triplets can break up
into a pair of spinons and thus will not remain sharp excitations.
To analyze the series for the triplet spectra, we use Euler transforms
and Pade approximants.
In Fig. 1, we show the Brillouin zone of the triangular lattice. In Fig. 2,
the excitation spectra for $\lambda=0.2$, $0.4$ and $1.0$ 
are shown along selected contours.
In Fig. 3, the spectral weights estimated by the [3/3] Pade
are shown along the same contours.
It is evident from these plots that as $\lambda$ is increased,
the triplet dispersion develops a sharp minimum at the
ordering wavevector of the triangular-lattice Heisenberg model.
The spectral weight associated with the triplets is
rapidly reduced over much of the Brillouin zone,
however, in the vicinity of the ordering wavevector, the
spectral-weights continue to increase with $\lambda$, and
the triplet excitations remain sharp. This provides further
evidence for the absence of an intermediate phase in this model,
and a direct transition from the local-singlet phase to the
magnetically ordered phase.

\begin{figure}
  \protect\centerline{\epsfig{file=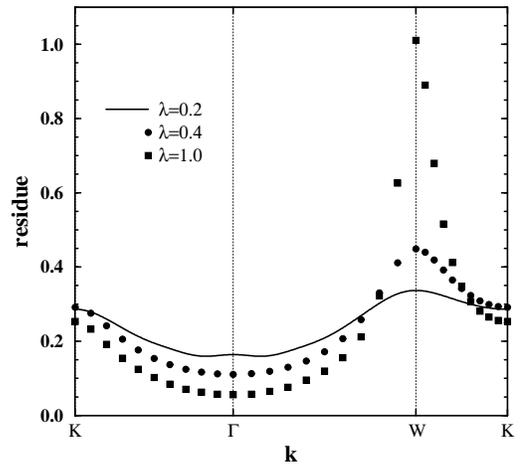, height=2.5in}}
  \caption{Spectral weights along selected contours}
\label{fig:residue}
\end{figure}

These results provide further confirmation that the 
single-plane triangular-lattice
antiferromagnet is ordered.\cite{triangle0}
However, the ratio of the ferromagnetic to antiferromagnetic
critical points is almost an order of magnitude smaller for the
triangular lattice than for the square-lattice.
This, together with previous
perturbative studies of the triangular lattice Heisenberg
model\cite{SH,ESY} 
shows that the antiferromagnetic ordering in the triangular-lattice 
model is much less robust.

In conclusion, in this paper we have studied 
the quantum phase transitions in the
bilayer triangular-lattice Heisenberg models in a strong coupling
expansion. For ferromagnetic intralayer coupling, the transition to
the ordered phase is found to be in the $3D$ classical
Heisenberg universality class. The antiferromagnetic intraplane
coupling case appears to be quite different. We find evidence that
there is a transition to an ordered phase at much larger values of
$\lambda$ and the transition is in the universality class of the
stacked triangular lattice.  This, together with the result that the
triplet spectral weight near the ordering wavevector continues to grow
with $\lambda$ suggest that in this model, a phase with free spinons
does not exist and there is a direct transition from the local singlet
phase to the 3-sublattice ordered phase.  This study lends further
support to the idea that the single-plane spin-half triangular-lattice
Heisenberg model is ordered, but that this ordering is much less
robust than for the square-lattice.

Acknowledgements: 
We would like to thank Subir Sachdev and Oleg Starykh for valuable discussions.
This work is supported in part by
the US National Science Foundation under Grants No.
DMR-96-16574.

\end{document}